\documentclass[10pt, conference]{IEEEtran}
\IEEEoverridecommandlockouts
\usepackage{cite}
\usepackage{amsmath,amssymb,amsfonts}
\usepackage{algorithmic}
\usepackage{graphicx}
\usepackage{dblfloatfix}
\usepackage[table]{xcolor}
\usepackage{textcomp}
\usepackage{url}
\usepackage[section]{placeins}
\def\BibTeX{{\rm B\kern-.05em{\sc i\kern-.025em b}\kern-.08em
    T\kern-.1667em\lower.7ex\hbox{E}\kern-.125emX}}
\begin{document}

\title{FuzzerGym: A Competitive Framework for Fuzzing and Learning\\
}

\graphicspath{ {C:\Users\bdrozd\Documents\RIOT-Generalization\Fuzzer\All_Graphs} }

\author{\IEEEauthorblockN{William Drozd}
\IEEEauthorblockA{National Robotics Engineering Center\\
Carnegie Mellon University\\
Pittsburgh, PA 15201\\
bdrozd@rec.ri.cmu.edu
}
\and
\IEEEauthorblockN{Michael D. Wagner}
\IEEEauthorblockA{\IEEEauthorblockA{National Robotics Engineering Center\\
Carnegie Mellon University\\
Pittsburgh, PA 15201\\
mwagner@cmu.edu}}}

\maketitle

\begin{abstract}
Fuzzing is a commonly used technique designed to test software by automatically crafting program inputs. Currently, the most successful fuzzing algorithms emphasize simple, low-overhead strategies with the ability to efficiently monitor program state during execution. Through compile-time instrumentation, these approaches have access to numerous aspects of program state including coverage, data flow, and heterogeneous fault detection and classification.   However, existing approaches utilize blind random mutation strategies when generating test inputs. We present a different approach that uses this state information to optimize mutation operators using reinforcement learning (RL). By integrating OpenAI Gym with libFuzzer we are able to simultaneously leverage advancements in reinforcement learning as well as fuzzing to achieve deeper coverage across several varied benchmarks.  Our technique connects the rich, efficient program monitors provided by LLVM Santizers with a deep neural net to learn mutation selection strategies directly from the input data. The cross-language, asynchronous architecture we developed enables us to apply any OpenAI Gym compatible deep reinforcement learning algorithm to any fuzzing problem with minimal slowdown.
\end{abstract}

\section{Introduction}
Fuzzing is a widely used testing technique that automatically probes software programs by repeatedly causing them to execute generated inputs.  These inputs are designed to find vulnerabilities in a target program by exercising as many code paths as possible under varying conditions.  Approaches to generating these sets of program inputs (called an input corpus) vary due to the trade-off between analysis and test execution speed.  Black-box fuzzing techniques \cite{b10}, \cite{b20}, \cite{b26} use the least amount of program information enabling rapid generation of inputs, with the consequence that they produce the least depth of code coverage. White-box fuzzing \cite{b12} performs detailed analysis of source-code uncovering deep paths through program branches using constraint solvers. While white-box fuzzers can achieve high code coverage, but require the most computational cost for each generated input.  Greybox techniques find a balance between these two extremes emphasizing simple input generation strategies with the ability to efficiently monitor program state with minimal overhead \cite{b40}, \cite{b43}.  Greybox fuzzers such as libFuzzer \cite{b24} and AFL \cite{b47} instrument programs during compilation to effectively monitor progress during testing,  enabling them use the amount of coverage achieved so far to dynamically guide input generation for future executions. These Coverage-based Greybox Fuzzers (CGFs) maintain a corpus of effective test inputs which are repeatedly, randomly modified before being executed using a set of simple data manipulation operations called mutators. Each time a mutated input covers a new section of code, the input is added to the corpus so that it can be selected again for further mutation.  By dynamically building the corpus in this way, CGFs can incrementally build a repository of useful inputs to guide deeper exploration.  In practice, collaborative learning between fuzzing researchers occurs indirectly by sharing these corpuses of “good” inputs, so that future tests can start from inputs structured appropriately for a given program.  

We observe that the CGF fuzzing loop is a good architecture for integrating automated learning \cite{b32} because tests are often executed very quickly (as fast as a tested function allows) with rich program state provided by compiler instrumentation \cite{b40},\cite{b43}. Despite the fact that an average 24 hour fuzzing run may execute billions of mutations, research attempting to improve this process is sparse, with current state-of-the-art being uniform random selection.  While uniform randomness combined with many executions may eventually produce a good result, it is common practice to periodically restart runs as they get stuck in coverage local minimum \cite{b24}. We propose to apply reinforcement learning to the fuzzing process, to more intelligently select mutation operators (Actions) to achieve deeper code coverage (higher Reward) in less time. By improving mutation operator selection, our approach is able to more effectively drive the local search from existing seeds to interesting new test inputs.

Although the ubiquitous "fuzzing loop" (Fig. 1) bears striking similarities to the state-action-reward sequence in the highly active field of reinforcement learning (RL) \cite{b6}, \cite{b36}, there are substantial challenges preventing its direct application to fuzzer mutator selection.

\begin{figure*}[t!]
    \centering
    \includegraphics[width=\textwidth]{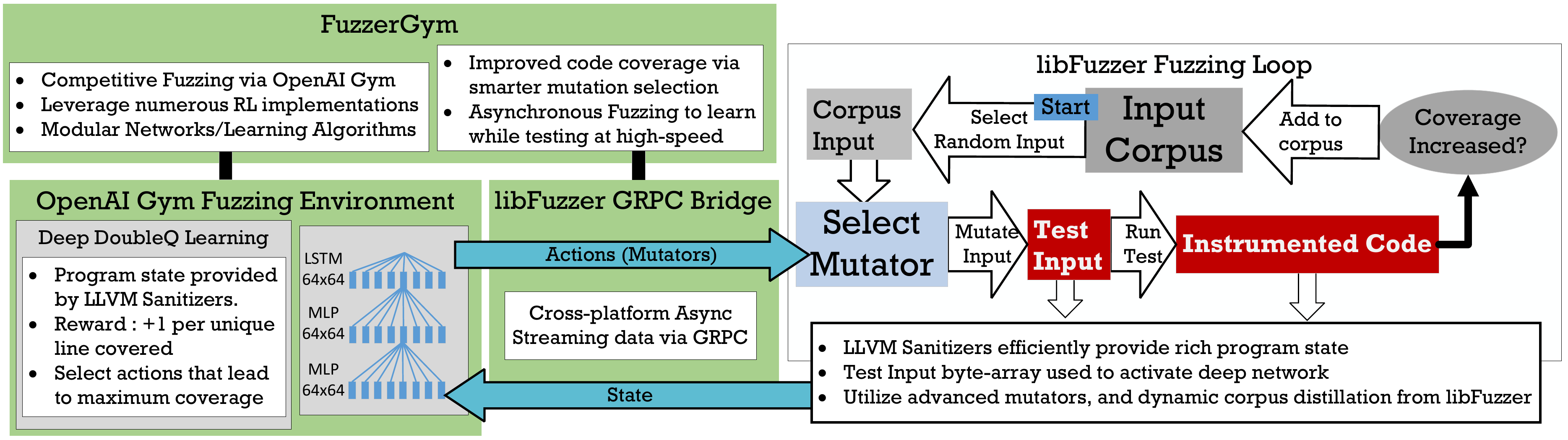}
    \caption{Fuzzer Gym High-level Architecture}
\end{figure*}

Manually engineering features for this learning process is infeasible due to the huge state space of possible program inputs, with potential difficulties generalizing these features to new programs. We demonstrate that like other aspects of fuzzing, mutation operator selection can benefit from automated feature engineering via deep learning techniques\cite{b32}. By learning a deep representation of input data we are able to produce a model that can achieve greater coverage than libFuzzer not only on the trained program but across a wide variety of other programs without retraining.

Despite advances in deep reinforcement learning, there remains a huge disparity between the execution time of deep RL networks (ms) and the potentially simple functions being tested by the fuzzer ($\mu$s).  To overcome this, we design an asynchronous architecture for fuzzing while reinforcement learning that treats this problem as a real-time Atari game (a video game console developed in the late 1970s whose simple games are commonly used to benchmark RL algorithms \cite{b27}) instead of a discrete sequential decision problem. By integrating our architecture with the OpenAI Gym RL framework and the widely used CGF libFuzzer \cite{b24} we can immediately obtain a large library of fuzzing benchmarks \cite{b25}, tools for large scale fuzzing deployment \cite{b15},\cite{b39} and cutting edge Deep RL algorithms \cite{b3}, \cite{b27}. 

Summary of our main contributions:
\begin{itemize}
\item We show that by using reinforcement learning we can train a fuzzer to select mutators in an effective way, without slowing down fuzzer performance.
\item By using an asynchronous frame-skipping technique originally developed to improve Atari game-playing, our RL agent is able to overcome the disparity in execution times between a fuzzer ($\mu$s) and a deep neural net (ms) \cite{b3}    
\item Our approach is capable of achieving superior average line coverage on a majority of benchmarks (4 out of 5)
\item We show that our RL-Fuzzer is able to penetrate more deeply into several different types of software even when the baseline \cite{b24} is given substantially more runs.
\end{itemize}

\section{Background and Motivation}
Despite their successes, grey-box fuzzers tend to require a huge number (on the order of millions of executions even for simple program) of test invocations (and mutations) to achieve relatively deep code coverage. This is primarily because their success is highly dependent on the ability of the fuzzer to randomly generate inputs that pass various conditional checks leading to deeper code paths. One solution is to deploy many fuzzing runs simultaneously \cite{b39} across many different machines. Large scale fuzzing deployments have found over 2000 bugs across more than sixty projects running at scale in many of the largest software companies. This process is run continuously \cite{b15}, \cite{b39}, with corpuses being shared among researchers to collectively improve future runs. 

This focus on corpus generation and selection is also indicative of the majority of current research. Skyfire \cite{b45} builds structured corpus inputs more effectively by using a probabilistic context-sensitive grammar. Vuzzer \cite{b33} uses program state information to dynamically favor selection of inputs that lead to promising code paths, while Learn\&Fuzz \cite{b13} uses statistical learning and neural nets to learns how to build structured inputs from a set of example test cases. Recently these approaches to structured input generation and filtering have expanded into deep learning techniques as well \cite{b13}, \cite{b29}, \cite{b32}.

While incrementally building corpuses for each program tested is useful, continuous fuzzing is still required to cope with changes in actively developed software. Although the default practice is intuitive, mutating unstructured data (starting fuzzing with an empty corpus) has the advantage of being unbiased w.r.t. the types of data passed to a function being tested. Furthermore, a program that has never been tested will also begin at this empty input corpus state. 

Since mutators depend on the input they are modifying, it follows that mutator selection can benefit from consideration of the structure of the input it is operating on. In particular, we show that making informed decisions about mutator selection based on the structure of the input data will lead to greater code coverage in a shorter amount of time.

\subsection{libFuzzer Mutation Operators}
The Fuzzing Loop (Fig. 1) used by libFuzzer is basically identical to those used by other mutational fuzzers. Execution begins with a ‘seed input’, a mutator is selected and applied (an action in our RL context), and a new input is created and tested. This new input might cause either a rewarding outcome (more coverage) or not (no new coverage). If the new input causes an increase in coverage, it is added to the input corpus to be selected in the future, otherwise it is discarded. 

The philosophy behind coverage-guided greybox fuzzers is that rather than spending enormous computation trying to understand the complex sequence of transfer functions that define a program, it is easier to rapidly try different inputs similar to previous inputs known to increase coverage. By repeating these random perturbations many times, mutational fuzzing will at some point create an input that increases coverage, thereby adding a new “valuable” input to the corpus which will be selected for mutations in the future (Fig. 1). As this process is repeated over time (called the “Fuzzing Loop”) what emerges is a corpus that covers a large number of lines of code without requiring any understanding of the format expected by the tested software. 

\setlength{\arrayrulewidth}{1.25pt}
\begin{table}[htbp]
\caption{Set of libFuzzer mutation operators}
\begin{center}
\rowcolors{2}{gray!25}{white}
\begin{tabular}{|p{1.8cm}|p{5.25cm}|}
\hline
\textbf{Mutator} & \textbf{Description} \\
\hline
EraseBytes & Reduce size by removing a random byte  \\
\hline
InsertByte & Increase size by one random byte \\
\hline
InsertRepeated Bytes & Increase size by adding at least 3 random bytes \\
\hline
ChangeBit & Flip a Random bit \\
\hline
ChangeByte & Replace byte with random one \\
\hline
ShuffleBytes & Randomly rearrange input bytes \\
\hline
ChangeASCII Integer & Find ASCII integer in data, perform random math ops and overwrite into input. \\
\hline
ChangeBinary Integer & Find Binary integer in data, perform random math ops and overwrite into input \\
\hline
CopyPart & Return part of the input \\
\hline
CrossOver & Recombine with random part of corpus/self\\
\hline
AddWordPersist AutoDict & Replace part of input with one that previously increased coverage (entire run) \\
\hline
AddWordTemp AutoDict & Replace part of the input with one that recently increased coverage \\
\hline
AddWord FromTORC & Replace part of input with a recently performed comparison \\
\hline
\end{tabular}
\label{tab1}
\end{center}
\end{table}
The random nature of mutators (Table 1) is what drives the powerful local search for useful inputs, but as a consequence, tests often need to be repeated many times to obtain additional coverage. While the selection of a new input from the corpus is always influenced by the preceding fuzzing steps, the actual relationship between an initial corpus and its contents later in a testing run are difficult to predict. This input selection process is made even more unpredictable due to the stochastic nature of the mutators used to produce new inputs. Furthermore, since the most successful fuzzers perform corpus distillation (an online technique for reducing corpuses to a minimal covering set), the distribution from which inputs are selected can change at any time in unpredictable ways. 
 
Unfortunately, in practice, the corpus is usually the only end product of a fuzzing run. The actual process of testing remains largely unobserved. We intend to expand the focus of fuzz testing beyond the corpus to the process that occurred to create it. Using our architecture combined with existing tools like OpenAI-Gym and Tensorflow, we can record the test-inputs (states) generated during testing and the actions that caused them. We show that by learning about mutation operator selection during fuzzing for a single program, that it is possible to transfer this knowledge to several other unrelated programs without additional training.

In addition to providing a platform for optimized mutator operator selection, it is also possible to use our RL framework to evaluate the efficacy of novel mutators or sets of mutators. Recent research towards more complex and powerful mutators FairFuzz \cite{b22} or generative techniques for mutator discovery \cite{b29} can benefit from an RL based evaluation because certain mutators may become more potent only in certain combinations or sequences, a discovery that would be difficult to uncover through exhaustively building mutator sets or parameterizations. For example if a long period has passed without achieving additional coverage, attempting to break through a blocking conditional (via the AddWordFromTORC operator) may lead to the greatest reward.

\begin{figure*}[t!]
    \centering
    \includegraphics[scale=1.0]{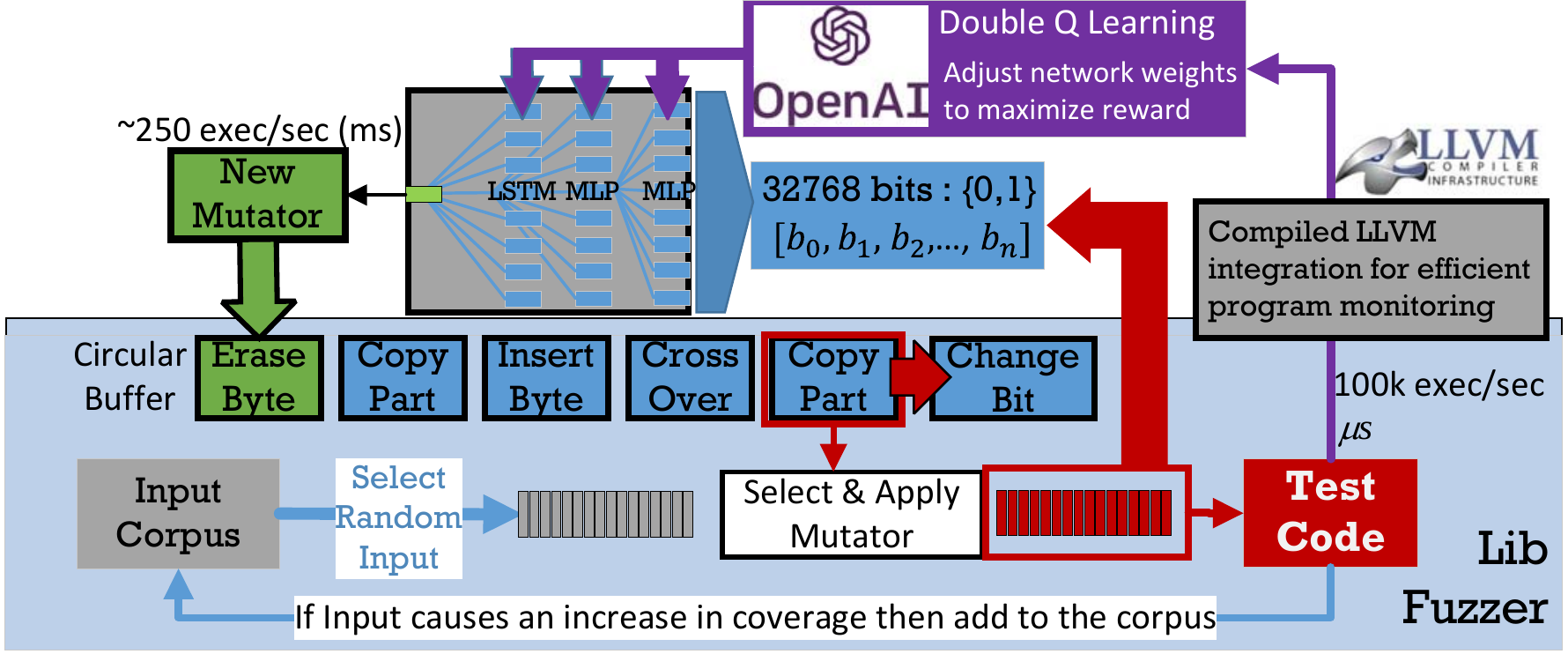}
    \caption{Asynchronous integration of OpenAI Gym and LLVM-libFuzzer}
\end{figure*}

\subsection{Efficient program state monitoring with LLVM}
A key differentiation between the three primary classes of fuzzer black-box, grey-box and white-box is the extent of program analysis that occurs during testing. Black-box fuzzing ignores state entirely to achieve maximum test throughput, whereas white-box techniques focus on heavyweight analysis to produce greater impact with each attempt. The difference between these two approaches highlights the trade-off inherent with all fuzz testing techniques: when is it worth it to spend time and effort on analysis? Our approach shows that by using state provided by compile time instrumentation, it is possible to learn how to fuzz more effectively with minimal overhead. 

This is achievable due to the tight integration of libFuzzer with the Sanitizers provided by the LLVM compiler \cite{b21} and a major reason why we chose to integrate with libFuzzer over other tools. As described in Table 2, LLVM provides five different types of runtime tools capable of detecting and classifying program faults as well as sophisticated data structures for dynamic code coverage tracking. Although the focus of our initial research effort is the creation of a fuzzer that can achieve superior line coverage, a major benefit of our framework is the ability to flexibly define RL reward functions using this rich Sanitizer state. This enables the creation of fuzzers targeting certain categories of bugs. All that is required to achieve this specialization is the creation of a reward function that targets the desired outcomes.

\begin{table}[htbp]
\caption{LLVM Sanitizers monitor coverage and provide fault classification and detection}
\begin{center}
\rowcolors{2}{gray!25}{white}
\begin{tabular}{|p{2.5cm}|p{5cm}|}
\hline
\textbf{Sanitizer Type} & \textbf{Description} \\
\hline
Thread Sanitizer & Data races, deadlocks \\
\hline
Address Sanitizer & Use-after-free, buffer-overflow, leaks. \\
\hline
Memory Sanitizer & Use of uninitialized memory \\
\hline
UndefinedBehavior Sanitizer & Detects undefined behavior \\
\hline
Coverage Sanitizer & Code coverage, execution frequency and caller-callee function relationships.\\
\hline
\end{tabular}
\label{tab2}
\end{center}
\end{table}

\subsection{Deep Reinforcement Learning}
Reinforcement Learning is a relatively simple concept, it assumes an agent is situated in an environment (fuzzer test program), makes observations about the current state (via LLVM) and uses this information to choose actions (mutation operators) that lead to the greatest reward (maximizing unique lines of code covered). RL has been used for years, but recent breakthroughs in deep learning have driven tremendous gains in RL performance across many highly challenging domains \cite{b19}. AlphaGo has defeated world champions in a complex sequential game \cite{b41}, breakthroughs in policy learning has taught robots to run \cite{b17} and Tensorflow how to play Atari \cite{b4}, \cite{b16}. The reason for these breakthroughs is the addition of deep neural networks to the RL tool set which enables automatic learning of features without human direction. While many RL algorithms are being concurrently developed across a diverse set of domains, in order to truly benefit from the research of others it is necessary to have a modular interface to the benchmark environments and the RL algorithms used to solve them. Fortunately, a set of standardized, reusable components for RL exists, provided by the widely used OpenAI Gym framework.

\section{Deep Reinforcement Learning to Fuzz}
By designing an OpenAI Gym environment that converts software testing into a competitive game, we are able to use state-of-the-art RL techniques to improve the mutation operator selection of libFuzzer across its many supported programs \cite{b25}. Although our initial experiments used only a single learning fuzzer, our architecture is designed to scale well in simultaneous fuzzing (ClusterFuzz \cite{b39}) and simultaneous RL learning (as in A3C \cite{b27}).

Perhaps the largest challenge combining reinforcement learning with fuzzing is the relative speed with which fuzzing execution occurs (Fig. 2). Greybox mutational fuzzer effectiveness is derived heavily from its execution speed which for the most part is only limited by the time it takes to execute the tested function. As discovered during our experimentation, testing could reach an execution speed of over 100k executions per second (\textasciitilde10$\mu$s per transition). These speeds were not momentary spikes, as we observed test runs of libpng that averaged 112295 executions per second sustained over a period of 400 seconds.  Although the average number of tests per second over a 24 hour fuzzing run is often much lower (10k executions per second), this entails approximately a billion tests over a 24 hour period. 

Another problem with combining learning and fuzzing is the enormous state space when using realistic test conditions. Given the libFuzzer default maximum input size of 4096 bytes, even a very simple mutator such as InsertByte can produce nearly $256^{4096}$ different outcomes without even considering variation in input structure. More complex operations that depend on the sequence of test invocations such as AddWordFromTempAutoDict, AddWordFromPersistAutoDict and AddWordFromTORC (Table 1) cause an even more explosive state space size which is dependent on the structure of the software being tested as well as the preceding sequence of fuzzing tests.  

We describe our approach to handle this complexity, via a novel asynchronous architecture combining reinforcement learning with fuzzing:
\begin{itemize}
\item We define the basic components of our RL problem formulation including State (Section III.A), Rewards (Section III.B) and our methodology for efficiently selecting Actions (Section III.C)
\item Utilizing inspiration from RL work in the Atari domain we design an asynchronous architecture for learning how to select mutators to increase coverage under realistic fuzzing conditions (Section III.C)
\item We describe our modular architecture designed to enable software researchers to use existing RL implementations with minimal prior knowledge. (Section IV)
\end{itemize}

\subsection{Modeling Program State in an RL Context}
To formulate an approach to solve any sequential decision making problem, it is necessary to determine how to represent a given position in the sequence, or in proper terminology its ‘state’. When possible it can be advantageous to structure state representations to suit a subset of programs being tested. For example in \cite{b13}, \cite{b45}, complex semantic structures are created by exploiting the grammars of PDF and XML files respectively. However, for our proposed architecture we operate directly on a byte array representation of the test input providing a universally applicable format which matches the input expected by both libFuzzer and the neural network based interface used by our RL framework.

While it is tempting to encode this data as-is using a sequence of numbers [0-255] corresponding to each value in the byte array, a ‘byte-level’ approach has a substantial drawback in that the test input data does not necessarily encode values directly (such as RGB value in an image) but may also encode bit-masks or other non-numerical values. We therefore, prefer the array-of-bits \cite{b32} technique to formulate test case inputs into our neural network. For all experiments, we fix the maximum input size to the default specified by libFuzzer (4096 bytes):

\begin{equation}
Test Input : {[0,255]}^{4096} \mapsto [0,1]^{32768}
\end{equation}

While our approach follows the technique described in \cite{b32} we did perform initial experiments using a byte representation and found empirically that the array-of-bits technique produced substantially better results for RL mutator selection.

\subsection{Flexible Fuzzing with RL Rewards}
A key advantage of using RL is the ability to specify a reward function to motivate fuzzing behavior that is perceived desirable. In our case, fuzzer ``high scores`` are strictly in terms of maximizing unique lines covered:

\begin{equation}
R_{t}=cov_{t} - cov_{t-1}
\end{equation}
Where:
\begin{itemize}
\item R\textsubscript{t} : Reward at time 't' 
\item cov\textsubscript{t}  : Number of unique lines covered at time 't' \\
\end{itemize}

The flexibility of the RL reward function means that it is possible to create fuzzers with various specific needs such as rapid bug finding or an increased reward for targeting specific code paths. Due to our integration with the LLVM Sanitizers (Table 2) it is straightforward to identify these other opportunities for additional rewards without additional instrumentation. This can lead not only to discoveries of useful testing strategies, but also to a more generalized understanding of the test inputs likely to elicit the desired reward response.
 
\subsection{Efficiently Selecting Mutator Actions}
Although mutational fuzzing contains a substantial number of actions with varying stochastic outcomes, the most challenging aspect of applying RL to mutator selection is the disparity between the amount of time it takes to evaluate a neural-net (milliseconds) and the time it takes to perform a single execution of a program being tested (microseconds). While it would be ideal from an RL learning perspective to treat this as a discrete fully observable problem (Markov Decision Process or MDP), it is our assessment this is impractical as it would require the fuzzer to wait for the RL evaluation before every test invocation. 

Since performing any test execution has at least some probability of achieving new coverage whereas performing no test does not, it makes little practical sense to force the fuzzer to be idle. Although it is not as difficult to devise an RL-Agent capable of outperforming random mutator selection on a per-action basis, it is rare that someone would use such a technique if it performs worse than random selection when both techniques are given equal testing time. 

We note that this disparity in execution time has been addressed previously in the Atari domains \cite{b3} via the creation of the Arcade Learning Environment (ALE).  If the game-emulator video frames are sampled periodically it is possible to learn while running the game at a faster rate. Frame-skipping \cite{b3} involves having the RL agent learn every ‘k’ frames (rather than every single frame) and repeating previous actions while waiting for new ones. This mitigates this disparity in execution time with a net positive effect to RL performance. In the case of the fuzzing domain, frame-skipping is even more critical to performance because executing far fewer tests puts a learning fuzzer at a dramatic disadvantage compared to the standard libFuzzer implementation. As illustrated in (Fig. 2) this can be done in an efficient way simply by maintaining a fixed size circular buffer of previous mutation actions.  

A consequence of the asynchronous nature of our architecture (which enables us to frame-skip while fuzzing) also means that we are solving a Partially Observable Markov Decision Process (POMDP) instead of the fully observable MDP that a serial discrete approach would provide. This means that instead of receiving every state input sent to the fuzzer, we are instead learning based on periodic observations of the true system state. This partial observability makes our problem formulation similar to the one discussed in \cite{b16}, who use an RL Agent that can play Atari games with a flickering screen. We mitigate this partial observability using the same approach described in \cite{b16}, which is to augment our network architecture with a Long Short Term Memory (LSTM) layer to ensure our RL agent can learn temporally dependent actions. We also note the success of using LSTMs on deep learning enhanced fuzzing via seed-filtering \cite{b32} as further motivation. 

While our asynchronous, partially-observable formulation does make the learning process more challenging, we believe that the trade-off greatly favors our asynchronous approach and is a more faithful representation of the real-world fuzzing RL problem.  
In summary, our approach differs from the serial MDP RL formulation in the following ways:
\begin{itemize}
\item \textbf{Partial Observability}: Our RL agent only receives samples of test inputs asynchronously to determine which mutators to choose. Although we cannot observe every state, we show that these snapshots provide an adequate approximation of the state to inform mutator selection
\item \textbf{Asynchronous Action Selection}: Since the rate of fuzz-testing far exceeds our RL network$'$s ability to generate actions, we instead replace the oldest selected action in a circular buffer as quickly as possible. Since libFuzzer can loop over this buffer continuously without blocking, we can test using our framework without a reduction in testing speed.
\end{itemize}

\section{Deep RL Fuzzing Benchmark Environment}
Although historically dominated by hand-engineered features, we demonstrate that like other aspects of fuzzing, mutation operator selection can also benefit from automated feature engineering via deep learning techniques \cite{b32}. 

To achieve this, we primarily utilized the RL learning framework Tensorforce \cite{b36} which provides numerous RL algorithms with an easy to use json-based configuration framework. Since Tensorforce is capable of learning any OpenAI compatible problem, we are able to use any of its existing features on our fuzzing problem. 

For our experiment, we chose to use Deep Double-Q learning (Section IV.A) with configuration parameters identical to those used for the ‘cart-pole’ benchmark provided by the Tensorforce software authors \cite{b35}. Although the complete configuration is available via the reference link, we note the use of prioritized rewards (50k capacity) in our configuration. Using this approach originally defined in \cite{b38}, past state-action-reward sequences are replayed to accelerate learning. These replayed memories are prioritized based on their estimated value to the learning process.

Since a key contribution of our work is a practical system that exceeds libFuzzer’s current capability, we use the entire list of 13 mutators (Table 1) as possible action choices for our RL agent. This is to ensure our RL agent is solving a problem comparable to the libFuzzer benchmark while also providing a test-bed that can in an emergent way provide information about which mutators are and are not useful for a given program. 

In this section we will describe in greater detail the three fundamental components necessary to apply the OpenAI Gym framework to fuzzing:

\begin{itemize}
    \item \textbf{RL Agent}: The implementation of the learning algorithm used to update the network weights to achieve maximum reward. (Double Deep Q Learning) (Section IV.A)
    \item \textbf{RL Network}: The underlying neural network architecture used to determine how an action is chosen for a given state. (LSTM based w/64 units) (Section IV.B)
    \item \textbf{Benchmark Environment}: The problem representation, which is the combination of Fuzzer configuration and program-under-test configuration. These are strictly versioned to ensure even minute changes to problem definitions are traceable and repeatable. (Section IV.C)
\end{itemize}

\subsection{Deep Double-Q Learning RL Agent for Mutator Selection}
The RL algorithm we used is based on Q-learning which means that it attempts to predict the utility of various actions without an explicit model --– in our case the mutator most likely to increase coverage (reward). This technique has been applied to many problems from elevator control to mobile robot navigation, and is commonly used as a benchmark in RL learning experiments \cite{b34}. Unfortunately, Q-learning regardless of whether it is instantiated using a tabular or multi-layered neural architecture (as in DQN) suffers from the problem of action value overestimation in stochastic domains. 

This problem is especially prevalent when learning to select mutation operators for fuzzing using our architecture because the asynchronous nature makes state observations and the subsequent reward arrivals noisy and potentially highly delayed. 

We therefore performed our experiments using a variant of Q-Learning, called Double Q-Learning which uses two sets of network weights to reduce overly optimistic value prediction. In our case it seemed to learn effectively even with a relatively short total training period (30,000 seconds). We define Double Q-Learning error following the definition of \cite{b14} which is a simple extension to the original DQN implementation designed to maximize computational efficiency:

\begin{equation}
\equiv R_{t+1} + \gamma Q ( S_{t+1}, argmax_{a}Q(S_{t+1},a;\theta_{t}); \theta_{t}')
\end{equation}
Where:
\begin{itemize}
\item R\textsubscript{t} : Reward at time 't' 
\item $\theta_{t}$  : Network Weights at time 't' \\
\end{itemize}

Using this algorithm the RL network weights are optimized by selecting the action `a` at timestep `t` that maximizes (argmax) the expected reward at the next time step, R\textsubscript{t+1}. 

Note that this formulation also considers the expected future reward by using a discount factor $\gamma \in [0,1]$ to determine how much the algorithm should value immediate reward relative to future predicted reward. For our research we leverage the fact that this approach was developed independently to be more successful on the challenging Atari domain without requiring us to implement it from scratch \cite{b35}. Although we primarily focus on a relatively simple RL learning configuration with a limited training time, it is our intent to use this as a starting point for more comprehensive benchmarking in future work.

\subsection{LSTM RL Network to Select Mutators with Memory}
As mentioned previously, to cope with the partial observability of our formulation, we augment our multi-layer perceptron based neural net with one that is able to learn over potential long sequences of states and actions. 

This Long-short Term Memory (LSTM) network layer \cite{b16} establishes a separate pathway for the flow of previous states and actions (memory) that enables remembering of useful experiences while forgetting ones that are no longer helpful. 

As illustrated in Fig. 2, this network architecture consumes bit-arrays and produces mutator selections while the DDQN algorithm (Section IV.A) optimizes weights to increase reward.

\subsection{Experimental Setup and RL Agent Training}

Once we have defined our network architecture and learning algorithm, it is necessary to orchestrate a training experiment to obtain network weights likely to produce effective action selections. Our goal is to train a fuzzer that can select mutation operators leading to the highest reward by maximizing total unique line coverage. During our research, we observed that unlike many typical RL problems, training fuzzers using short or even medium length runs tends to lead to RL Agents that are able to achieve the “low-hanging fruit” coverage most rapidly. For the goal of optimizing deep code coverage over a longer period this is undesirable as there is relatively little chance of breaking through a tough coverage barrier when each training episode is terminated quickly. 

Effectively covering low-hanging fruit lines of code while starting from an empty corpus usually empirically entails selecting size increasing mutators (such as InsertByte and InsertRepeatedBytes) over ones that try to create complex structure over a longer period of time (AddWordFromTORC, CrossOver, AddWordFromTemp, etc.). 

For libjpeg, we observed that the first coverage barrier (\textasciitilde327 lines covered) usually took 6000-9000 seconds of fuzzing (between 60M and 90M tests of the libjpeg decompress function) to break through. Although the actual amount of coverage obtained after penetrating this barrier may vary, this often corresponds to whether or not the fuzzer is able to create a valid jpeg header. It is our observation that training an RL Agent without experiencing these breakthroughs will result in a fuzzer learning to achieve easy coverage fast, without the ability to cover deeply.

In addition to these difficulties there are significant time pressures on the fuzzer configuration and setup. While the ideal approach in terms of performance may be to repeat these very long episode training sessions for many months, this is impractical in terms of computational requirements. 

Furthermore, delaying the start of a fuzzing experiment for a long period of time greatly reduces the ease of use of our framework. As eloquently described in the introduction of \cite{b46}, there is often urgent time pressure to start fuzz testing as soon as possible. It is for these reasons that we limit our training to very few episodes. For the network evaluated in Section V, we trained for only 3 episodes with each episode having a relatively long length of 10,000 seconds. This episode length was long enough to break through the 327 line barrier during training for 2 out of 3 episodes. 

We attribute our success in achieving deeper coverage at least partially to our ability to observe the patterns of inputs (states) likely to break through barriers and the mutators likely to cause them. While learning experiments are ongoing, we believe that being able to demonstrate effective learning with a very short total training time of about 8 hours (30k seconds total) is a promising start.

\section{Evaluation}
In this section we describe our experiment to evaluate the efficacy of our RL based fuzzer and the subsequent results. We compare our approach to libFuzzer on five benchmark tests found in the publicly available libFuzzer Test Suite \cite{b25}. While numerous fuzzing benchmarks exist, we chose the libFuzzer Test Suite due to its widespread usage \cite{b30} and the ease with which it integrates with our architecture. 

Our approach integrates with libFuzzer by making only one minor modification which is the source of the mutation operators. Since we only make a small modification to libFuzzer, we can reuse any of its existing features and specify libFuzzer parameters using definitions in our OpenAI Gym environment configuration. In order to achieve a fair comparison between our RL-Fuzzer and the libFuzzer benchmark we give each program the same time-limit per run of 24 hours. 

\begin{table}[htbp]
\caption{Summary of Benchmark Performance (lines covered)}
\begin{center}
\rowcolors{2}{gray!25}{white}
 \begin{tabular}{c|c|c|c|c|}
    \cline{2-5}
    \hline
    & \multicolumn{2}{c|}{libFuzzer} & \multicolumn{2}{c|}{Ours} \\
    \hline
    \multicolumn{1}{|c|}{Test Program} & Best & Avg & Best & Avg \\
    \hline
    \multicolumn{1}{|r|}{libjpeg} & 1049 & 402.8 & 1215 & 438.24 \\
    \multicolumn{1}{|r|}{libpng} & 561 & 396 & 582 & 427.6 \\
    \multicolumn{1}{|r|}{boringssl} & 876 & 813.48 & 895 & 811.4 \\
    \multicolumn{1}{|r|}{re2} & 2173 & 2114.6 & 2216 & 2182.4 \\
    \multicolumn{1}{|r|}{sqllite} & 905 & 857.96 & 2209 & 970.36 \\
    \hline
  \end{tabular}
\label{tab3}
\end{center}
\end{table}

\begin{figure*}[t!]
    \centering
    \includegraphics[width=.9\textwidth]{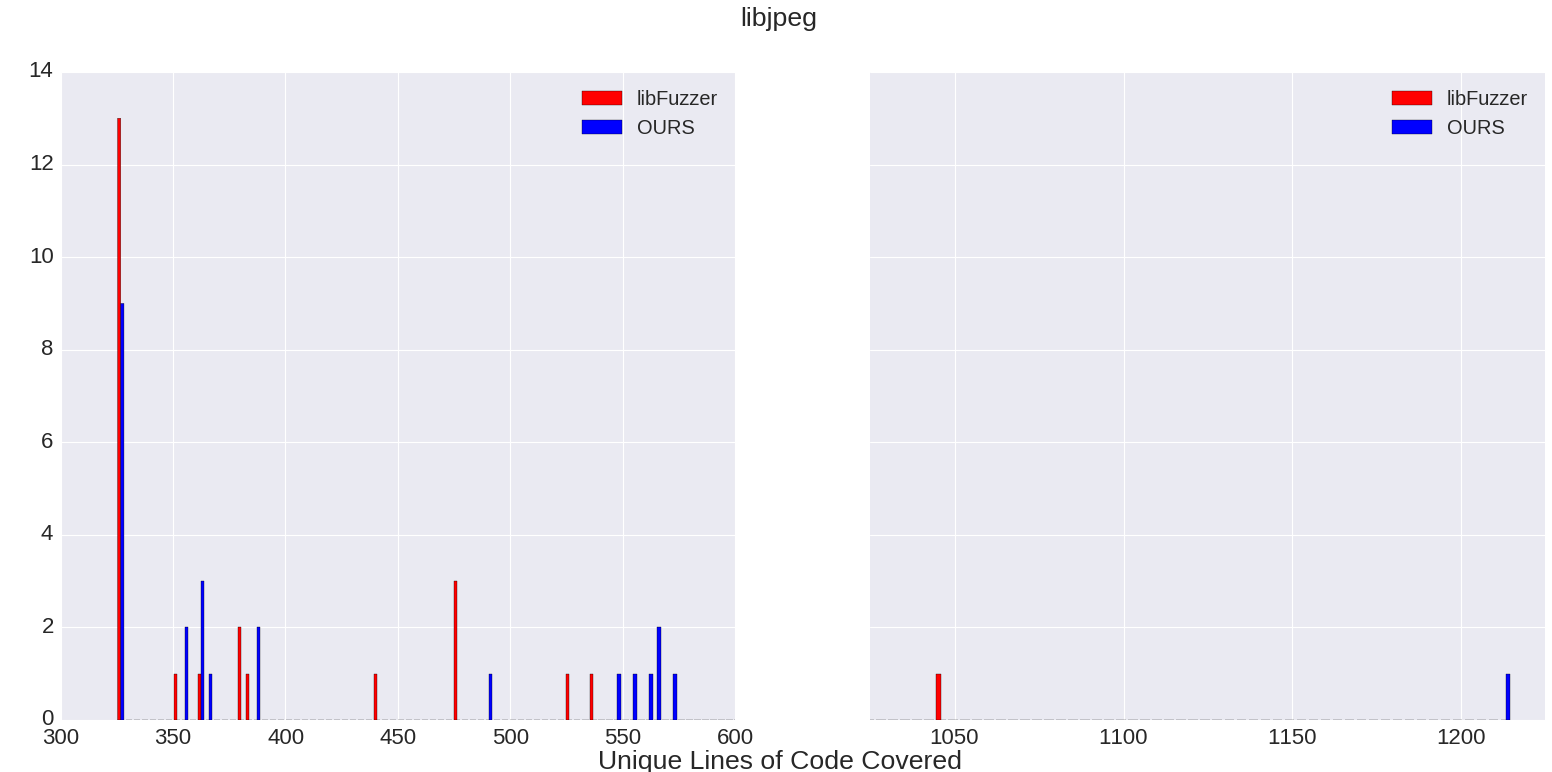}
    \caption{Total unique Lines covered for each libjpeg fuzzing run.}
    \label{fig3}
\end{figure*}

\begin{figure*}[b!]
    \centering
    \includegraphics[width=.9\textwidth]{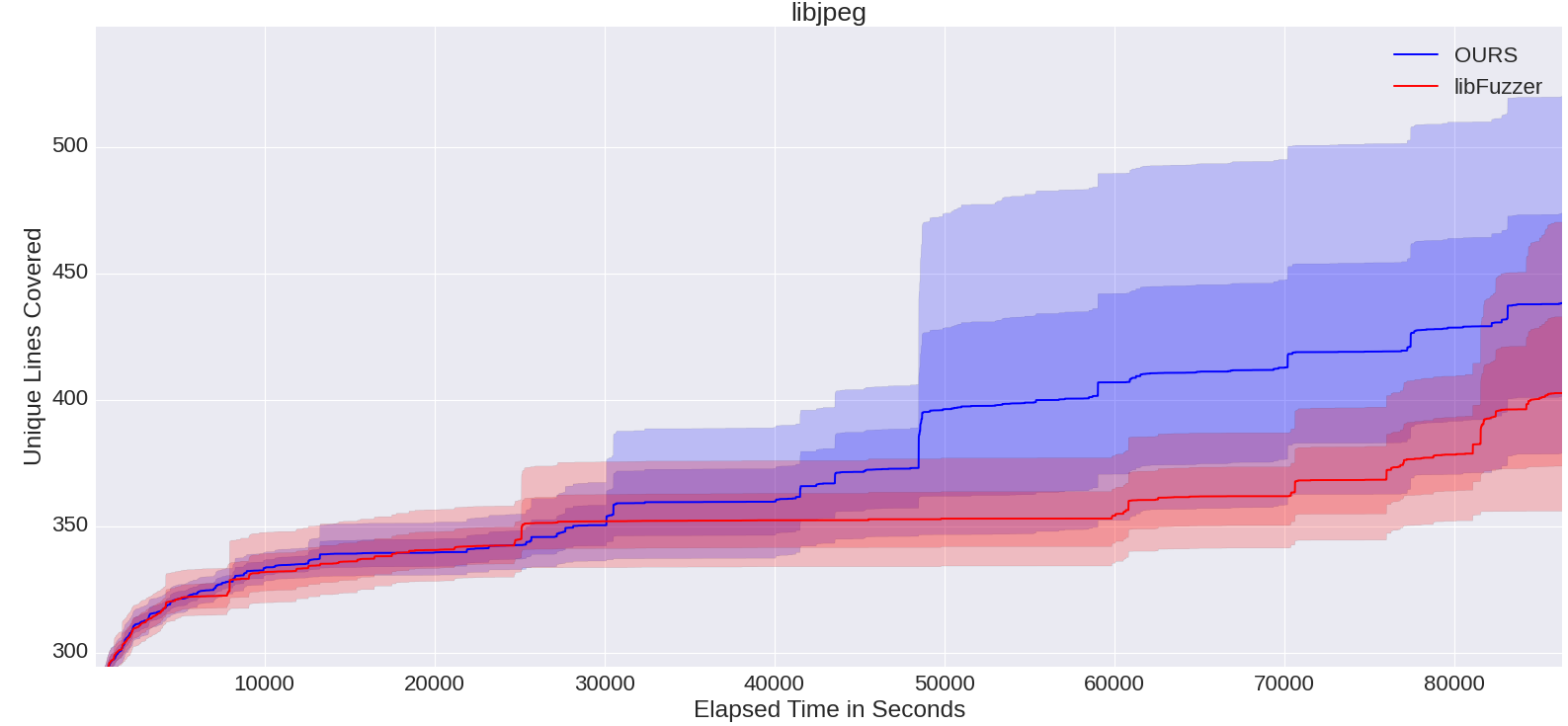}
    \caption{Average coverage of libjpeg over time with 95\% CI}
    \label{fig4}
\end{figure*}

Although recent research does not always repeat runs when evaluating fuzzers \cite{b5}, we instead follow the experiment design of \cite{b22} and repeat each run 25 times. For each experiment using the RL-Fuzzer we use fixed weights determined during the training phase described previously (Section IV.C). For each run we start with an empty seed corpus, which is automatically populated at initialization with one single-byte seed input prior to the start of fuzzing. We start from an empty to corpus in order to ensure unbiased fuzzing (thus maximizing the possibility of finding new paths through code) and to ensure we can evaluate differences in long-term strategy between the two fuzzing techniques.

Since coverage-guided greybox fuzzers add items only when coverage increases (Fig. 1), establishing a “useful” corpus early (when the corpus is expanded frequently and increasing coverage is easy) is an important factor to breaking through plateaus when coverage is hard. Our key results are as follows: (summarized in Table 3)
\begin{itemize}
    \item \textbf{Deepest Coverage}: We demonstrate that over all 25 runs, our approach is able to achieve the greatest depth of coverage on all 5 benchmarks.
    \item \textbf{Effective Average Coverage}: For (4/5) benchmarks we are able to show a substantial improvement in average lines of code covered.
    \item \textbf{Generalization}: Even though our RL agent was only trained on a single libjpeg function, it was able to generalize its effectiveness to several programs in substantially different domains.
\end{itemize}

\subsection{libjpeg - Decompress a JPEG}
First, we evaluate our RL Agent on the same program that it was trained on, which is an image decompression function from the libjpeg library. For all libFuzzer benchmarks the test function takes as input an array of bytes, so in this case libjpeg attempts to decompress these bytes into a valid JPEG file.

As discussed previously (Section IV.C) each training episode was run for 10k seconds, thus we used our fuzzer$'$s ability to break through the first coverage plateau as a proxy for learning how to fuzz in general.

As illustrated in (Fig. 3), we were able to penetrate the first coverage plateau (successfully decompressing the header at cov=327) about 50\% more often than libFuzzer. Our approach also seemed to be highly effective at passing the next coverage barrier located just below 500 lines of coverage.

\begin{figure*}[t!]
    \centering
    \includegraphics[width=.9\textwidth]{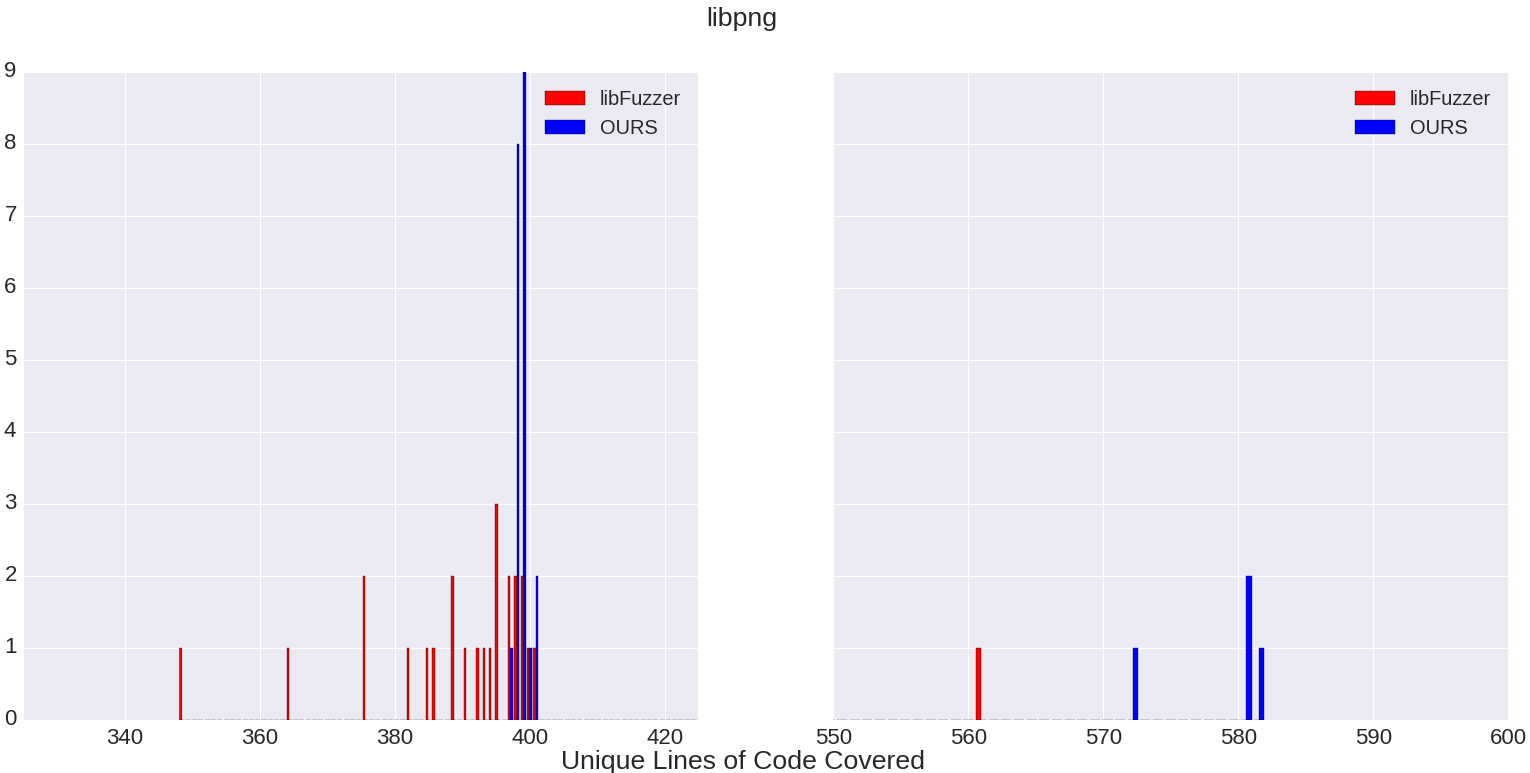}
    \caption{Total unique Lines covered for each libpng fuzzing run.}
    \label{fig5}
\end{figure*}

\begin{figure*}[b!]
    \centering
    \includegraphics[width=.9\textwidth]{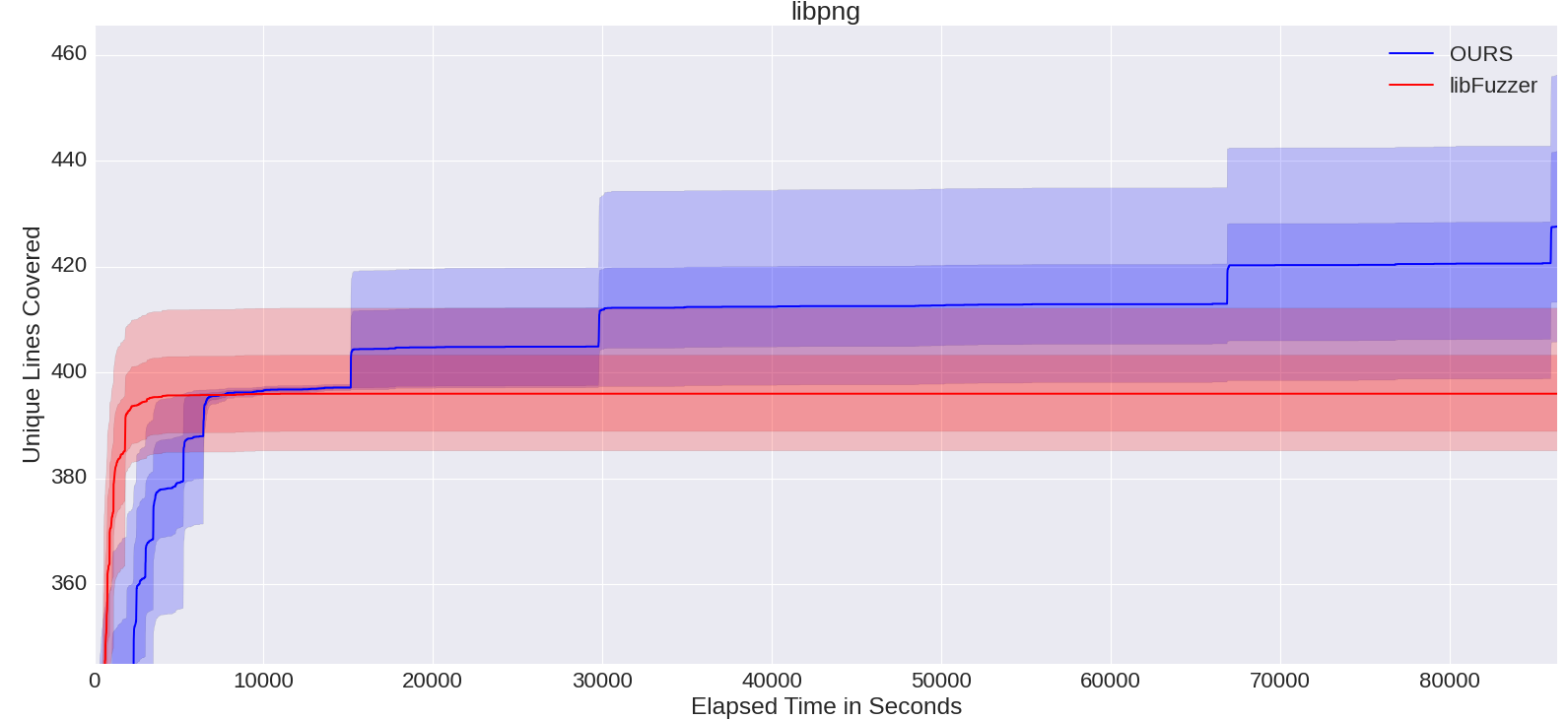}
    \caption{Average coverage of libpng over time with 95\% CI}
    \label{fig6}
\end{figure*}

While surpassing difficult barriers is important, we are primarily interested in the total coverage achieved after a reasonably long fuzzing run is completed. After 25 runs our approach is able to achieve deeper coverage than libFuzzer by 166 lines. Although it is the focus of our work to evaluate total coverage of each approach with a time limit of 24 hours, we also include a plot of the average coverage achieved by each fuzzer over time (Fig. 4).

Although it is helpful to consider coverage over time as one type of performance measure \cite{b5}, we note that this metric suffers from large spikes in standard deviation whenever a fuzzer breaks through one of these coverage plateaus. Although statistically this makes sense (as the run sometimes vastly exceeds the average coverage in a short period of time) it can be quite misleading as it would be impossible in our case for coverage to decrease.

Since one of the benefits of our approach is a single architecture for learning and benchmarking, we hope that by collecting many more runs it will be easier to understand how to further improve our ability to break through coverage barriers.

\begin{figure*}[t]
    \centering
    \includegraphics[width=.9\textwidth]{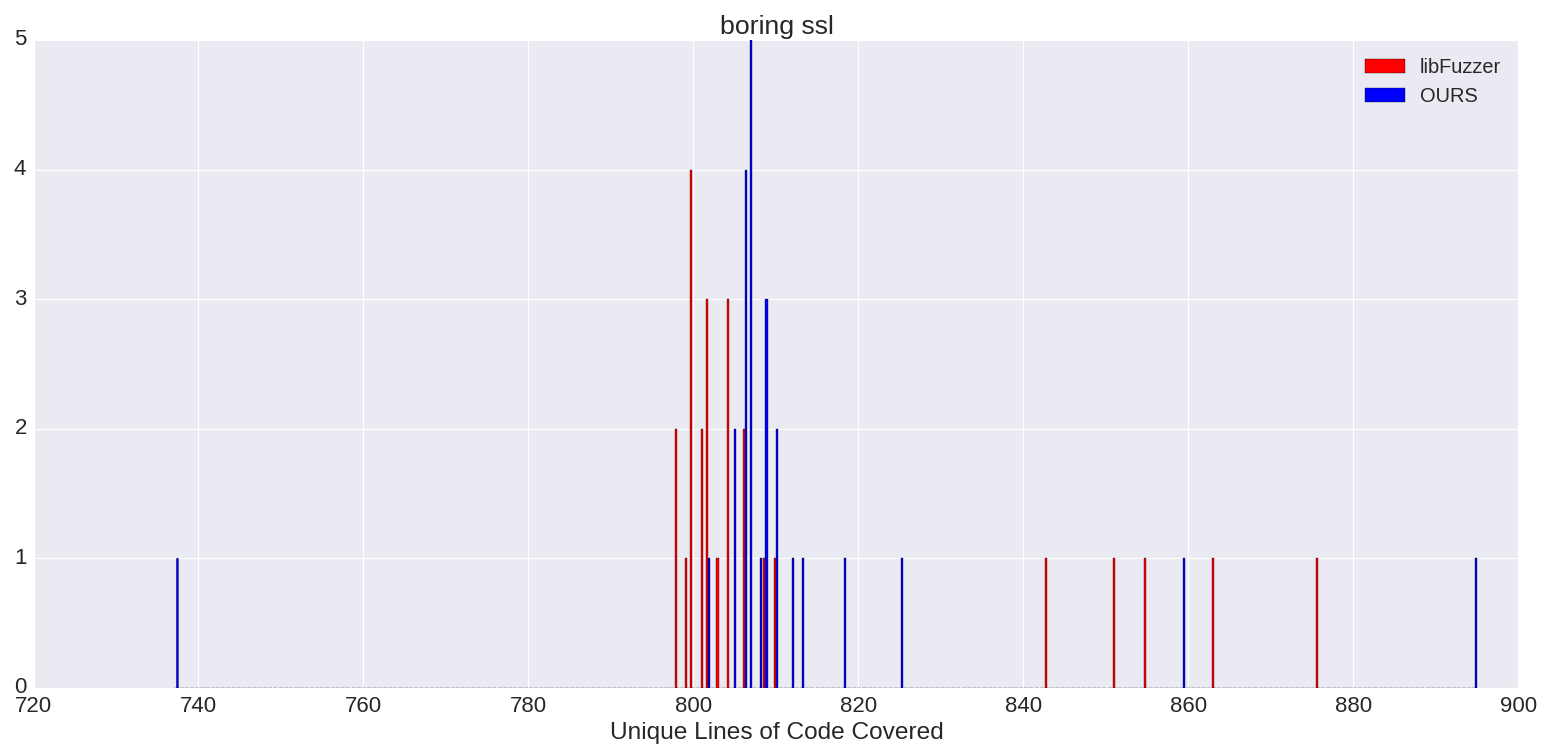}
    \caption{Total unique Lines covered for each boringssl fuzzing run.}
    \label{fig7}
\end{figure*}

\begin{figure*}[b!]
    \centering
    \includegraphics[width=.9\textwidth]{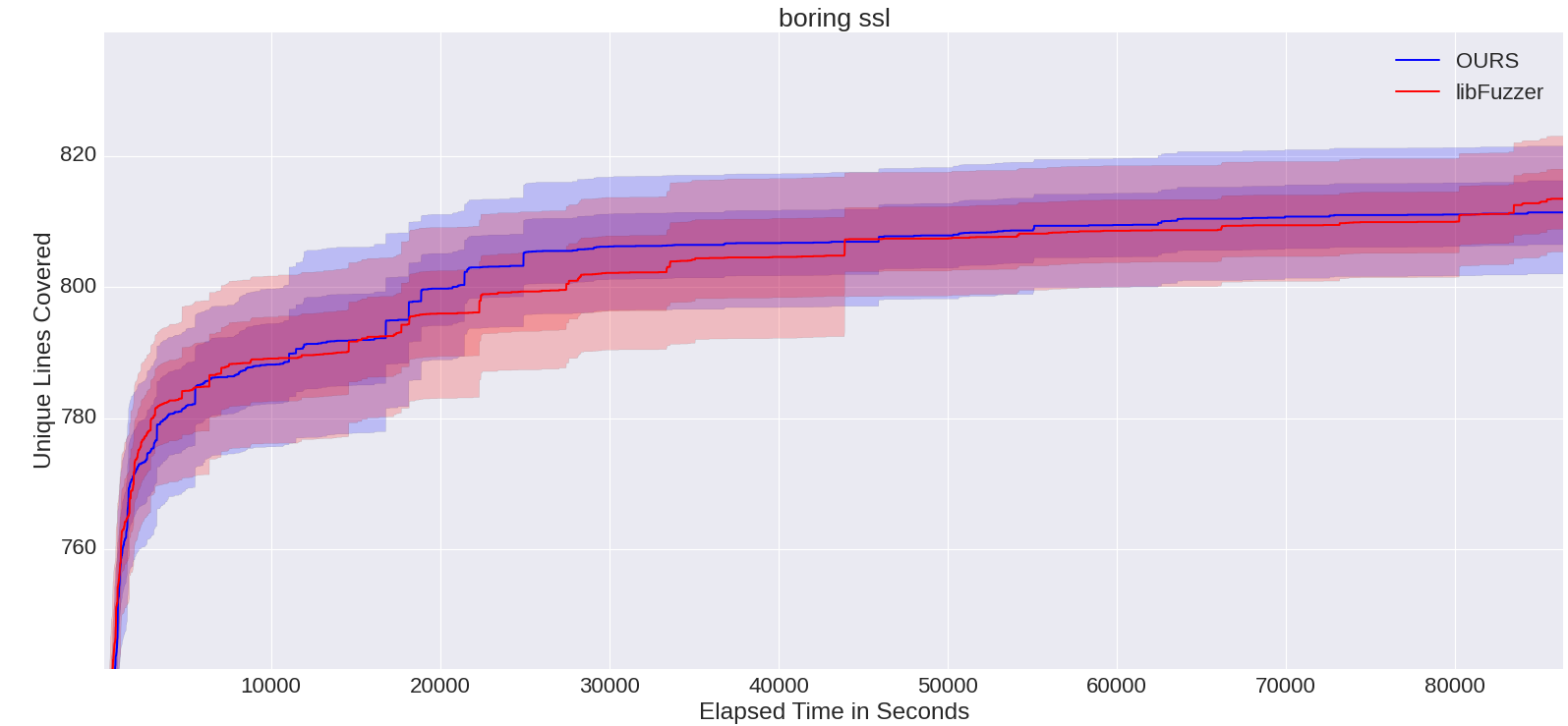}
    \caption{Average coverage of boringssl over time with 95\% CI}
    \label{fig8}
\end{figure*}

\subsection{libpng - Parse a PNG}
For our next experiment we evaluate our RL Agent on a similar test function from a different, previously unseen library --– libpng. Instead of generating bytes to be decompressed into a jpeg, we are instead constructing a png image file from the input byte array used by libFuzzer. 

This is an interesting benchmark as it will enable us to determine whether it is possible for a trained RL fuzzer to generalize to a similar type of program even if had never been encountered during training.

While our best run exceeds the coverage of libFuzzer by only 20 lines of code, as we will discuss in more detail in (Section V.F), libFuzzer seemed unable to exceed a coverage level of 570 lines even after a large number of additional trials (something our approach achieved 4 times as shown above in (Fig. 5) with scores of 572,581 ,581, and 582).
 
Similarly to libjpeg, our approach seemed to excel when passing early coverage barriers (this time in the $<$400 lines range). We also note a similar pattern to libjpeg when considering the average amount of coverage over time (Fig. 6) with large spikes in average coverage and standard deviation when our approach achieves the outlier breakthroughs at $>$570 lines of coverage.

\begin{figure*}[t]
    \centering
    \includegraphics[width=.95\textwidth]{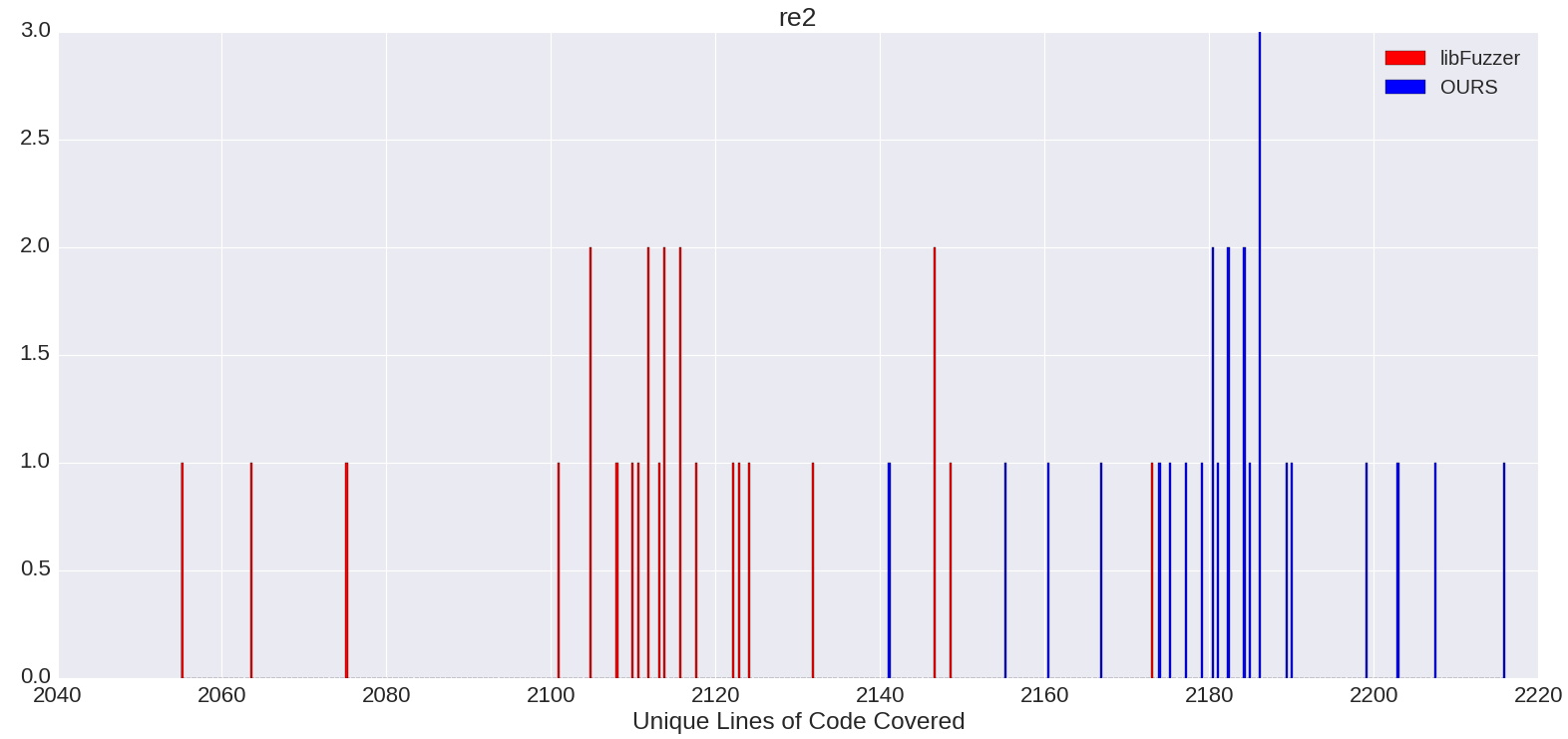}
    \caption{Total unique Lines covered for each re2 fuzzing run.}
    \label{fig9}
\end{figure*}

\begin{figure*}[b!]
    \centering
    \includegraphics[width=.95\textwidth]{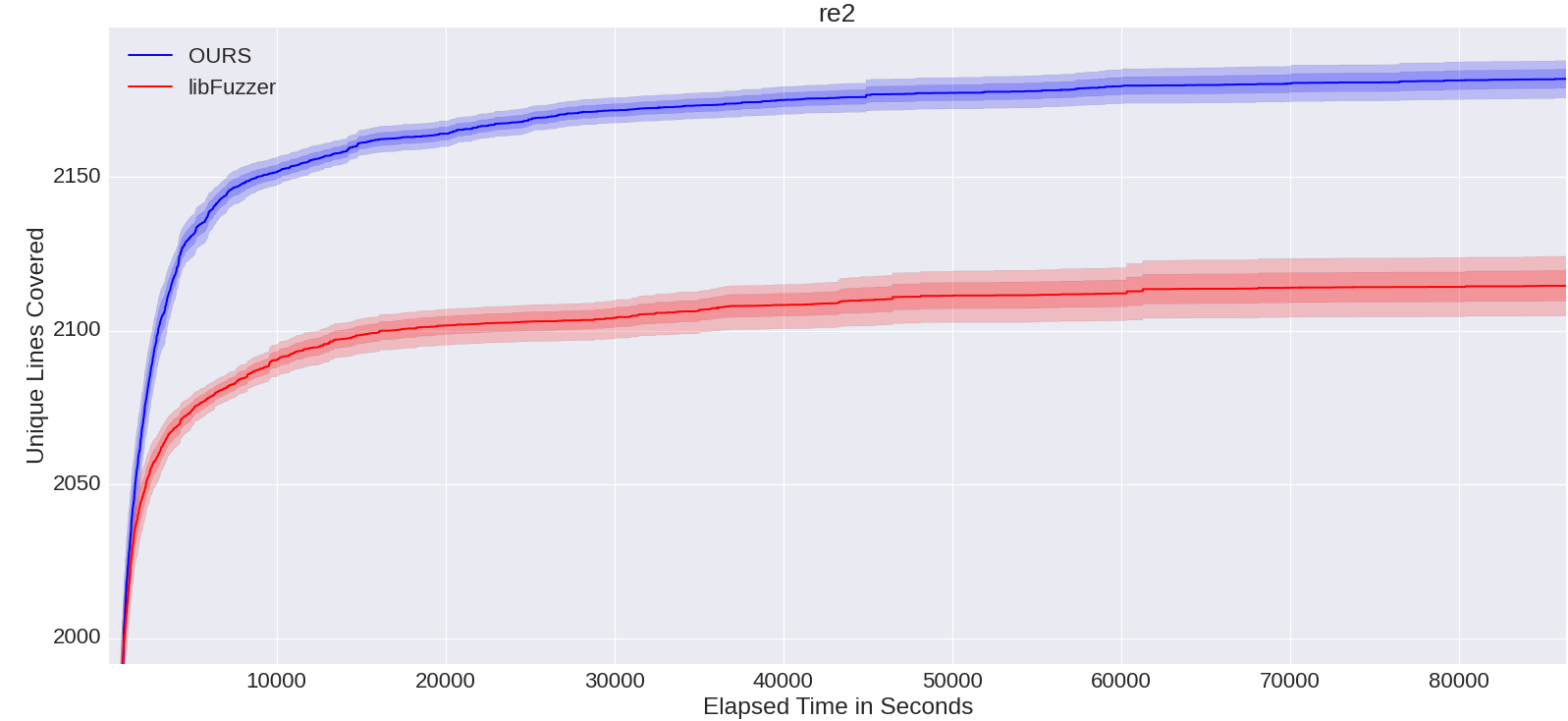}
    \caption{Average coverage of re2 over time with 95\% CI}
    \label{fig10}
\end{figure*}

\subsection{BoringSSL - Encode a Private SSL Key}
For our next experiment we evaluate our RL Agent on software that was quite dissimilar to the one encountered during training. This test function, to encode a key using the BoringSSL function $‘d2iAutoPrivateKey’$ showed the least improvement over the libFuzzer baseline of the five benchmarks evaluated. 

As with the other benchmarks we were able to achieve deeper coverage (895 vs. 876) than libFuzzer. While our approach seemed to do a bit better job penetrating the first coverage plateau at around 800 lines of coverage, (Fig. 7) libFuzzer seemed to do a better job penetrating the later plateau located around cov=840. 

Boring-SSL was the only benchmark tested where we achieved worse average coverage compared to libFuzzer (Fig. 8) although their scores using this metric are similar.

\subsection{re2 - Evaluate a Regular Expression}
We executed our RL Agent on another dissimilar test function that was not encountered during training, this time we evaluated each approach according to their ability to cover a regular expression library ‘re2’.

\begin{figure*}[t]
    \centering
    \includegraphics[width=.9\textwidth]{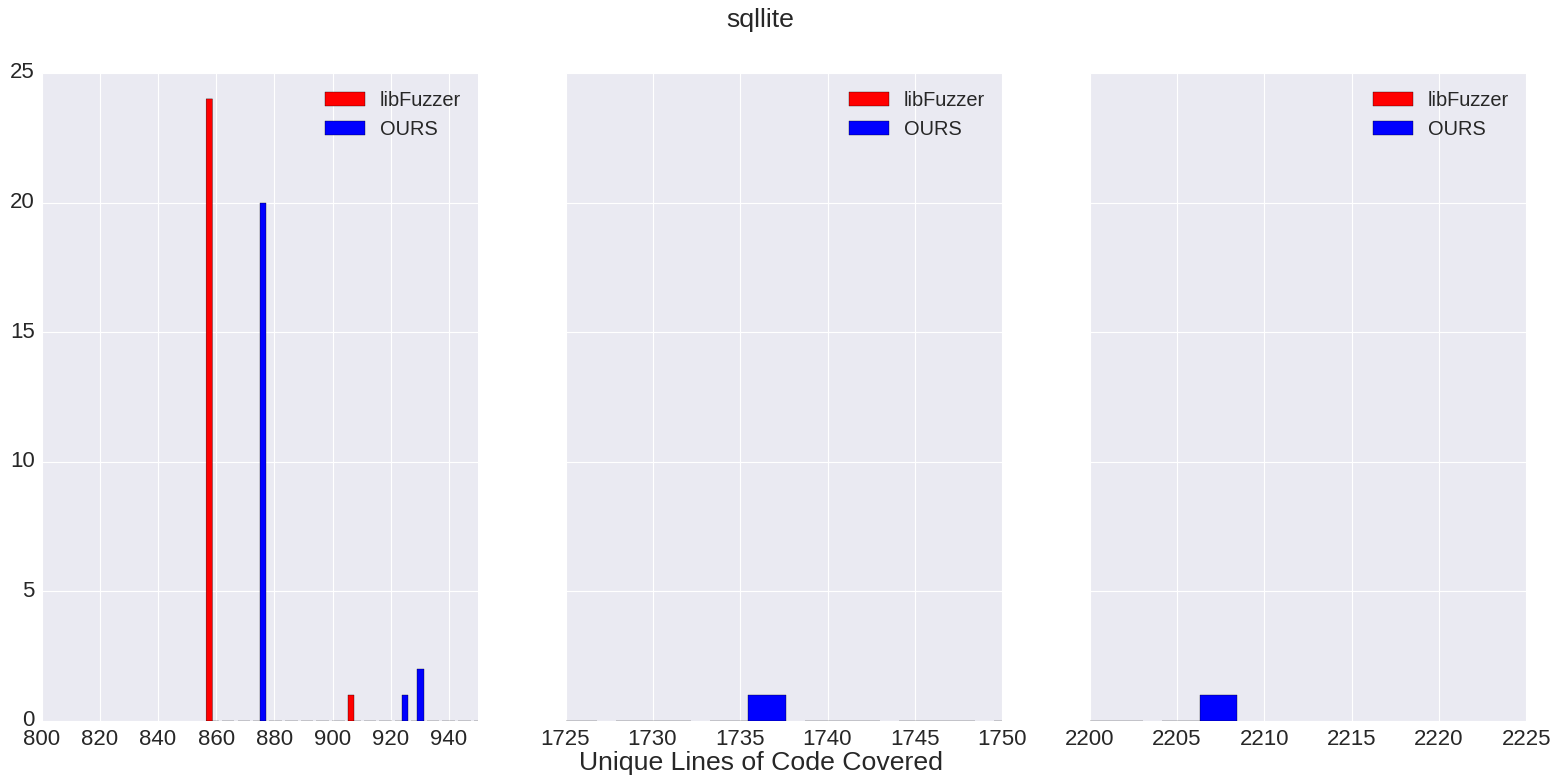}
    \caption{Total unique Lines covered for each sqllite fuzzing run.}
    \label{fig11}
\end{figure*}

\begin{figure*}[b!]
    \centering
    \includegraphics[width=.9\textwidth]{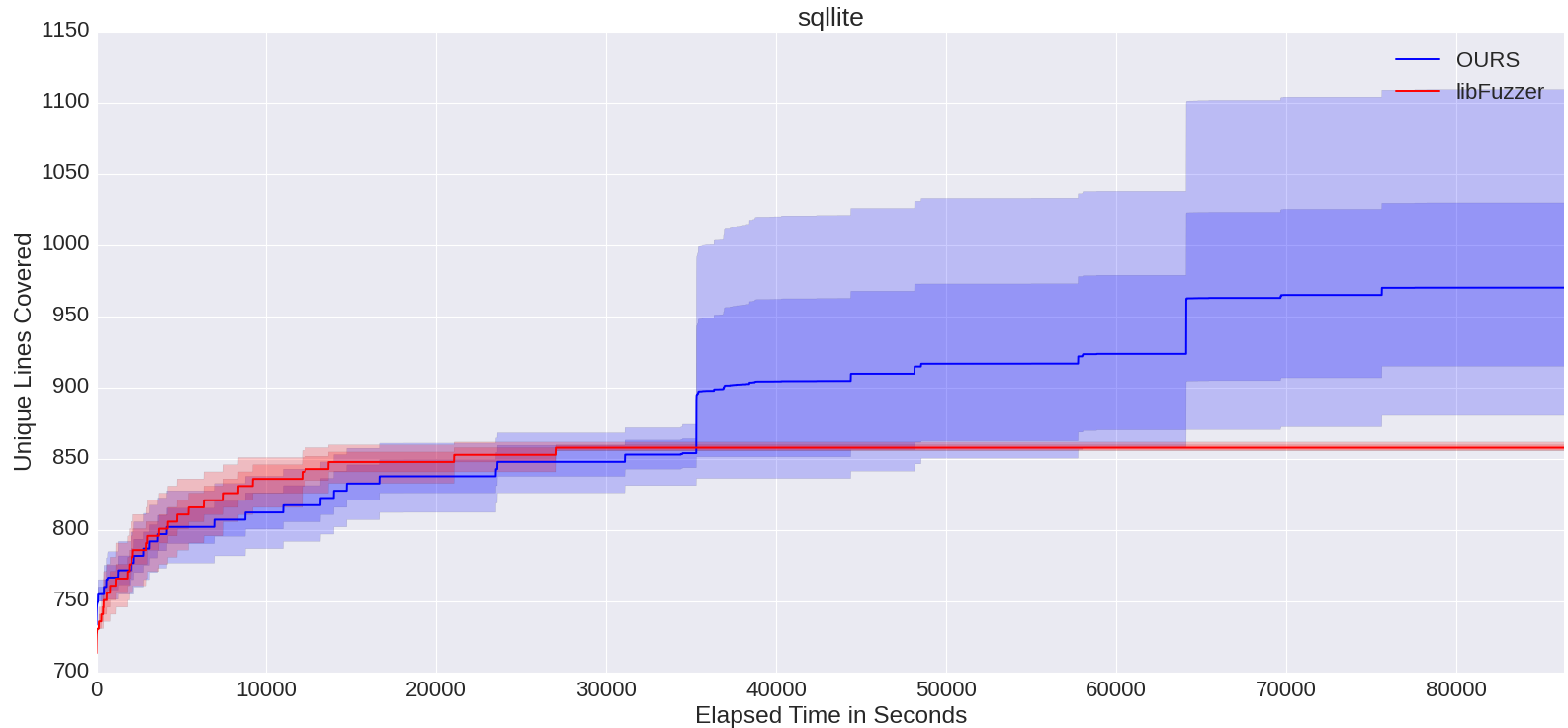}
    \caption{Average coverage of sqllite over time with 95\% CI}
    \label{fig12}
\end{figure*}

For this benchmark, the test function was RE2::FullMatch which runs a regular expression parser on an input. Unlike the other software evaluated RE2 was unique in that it was very easy to achieve coverage of a large number of lines of code without much effort. 

As with all other benchmarks, our approach was able to achieve a higher maximum coverage than libFuzzer ( Fig. 9), however the most striking difference is that we were able to achieve a higher amount of coverage than libFuzzer’s all-time best (2173) about 88\% (22/25) of the time. 

As shown in more detail in the coverage over time plot depicted in Fig. 10, both approaches were able to obtain 2000 lines of coverage after about 1000 seconds of execution. While the overall range of coverage obtained did not vary by a large amount there is a clear statistical advantage to using our RL based approach over libFuzzer.

\subsection{SQL Lite - Execute a SQL Query}
The final benchmark program was the popular database implementation, SQLLite. In this case the test function was ‘sqlifte3exec’ which is used to evaluate a SQL query. This benchmark is also substantially different from the training program, libjpeg. 

Of all software tested libFuzzer seemed to have enormous difficulty penetrating the first coverage barrier located at 856 lines of coverage. The best run produced by libFuzzer (cov=905) was the only run of all 25 that did not get stuck at (cov=856). This seems to indicate that our RL-Fuzzer has learned what we were trying to teach it which was a strategy for action selection that is more adept at passing early code barriers.

Since this first barrier had a huge impact in the case of SQLLite$'$s sqlite3exec function, the performance differences between the two approaches was dramatic (Fig. 11).

The graph showing code coverage over time predictably shows a huge spike in standard deviation corresponding to the massive breakthroughs of our two best runs (1735, 2209). However in this case our performance was so much better than libFuzzer, the average coverage was also statistically significantly better.

\subsection{Longer Fuzzing with libFuzzer doesn't help}
While we have demonstrated that our approach was able to achieve deeper coverage than libFuzzer over 25 runs, we collected some additional runs for each of our benchmark programs. 

As shown in Table 4, even after substantially more runs libFuzzer was not able to produce a level of coverage exceeding our approach. Although it would be interesting to determine exactly the number of runs necessary for libFuzzer to exceed our level of coverage, we eventually had to ‘give up’ on this pursuit due to time constraints of our own. We believe that a major contribution of our research is our ability to use this large scale deployment of fuzzers to train future RL models.
\begin{table}[htbp]
\caption{Even with numerous additional runs libFuzzer is not able to surpass our max depth of coverage on any benchmark}
\begin{center}
\rowcolors{2}{gray!25}{white}
 \begin{tabular}{c|c|c|c|c|}
    \cline{2-5}
    \hline
    & \multicolumn{2}{c|}{libFuzzer} & \multicolumn{2}{c|}{Ours} \\
    \hline
    \multicolumn{1}{|c|}{Test Program} & Best & Runs & Best & Runs \\
    \hline
    \multicolumn{1}{|r|}{libjpeg} & 1049 & 127 & 1215 & 25 \\
    \multicolumn{1}{|r|}{libpng} & 564 & 772 & 582 & 25 \\
    \multicolumn{1}{|r|}{boringssl} & 876 & 30 & 895 & 25 \\
    \multicolumn{1}{|r|}{re2} & 2175 & 71 & 2216 & 25 \\
    \multicolumn{1}{|r|}{sqllite} & 905 & 47 & 2209 & 25 \\
    \hline
  \end{tabular}
\label{tabsum}
\end{center}
\end{table}

\section{Related Work}
While other researchers have previously formulated fuzzing as a discrete Markov Decision process, there are substantial issues scaling their techniques to a real-world fuzzing context. Most notably the recent research \cite{b7} attempts to solve the same problem of mutator selection using RL. While both our research and the work presented in \cite{b7} has been undertaken independently and concurrently, the differences between our two approaches are numerous. Our work was designed with the ability to be practically integrated into large scale fuzzing (via libFuzzer) and reinforcement learning (via OpenAI) architectures, while the work in \cite{b7} is evaluated using a manually constructed plugin designed to work on a single program. 

Furthermore, the work in \cite{b7} despite being a small study already shows difficulty scaling to realistic fuzzing scenarios primarily for the reasons we detail in (Section III.C). Lastly and most importantly, we do not believe evaluating an RL based fuzzer using runs with a fixed number of actions is fair or practical as a true benchmark must account for the trade-off between speed and complexity. As the majority of fuzzing research indicates, this entails giving each approach equal time to perform testing rather than an equal number of actions. 

Other researchers have previously explored MDP based approaches for improved fuzzing, most notably in \cite{b46} where the authors use a Multi-armed Bandit approach to select (program, seed) pairs most likely to lead to the discovery of a bug. By slicing the fuzzing run into discrete epochs of 10s they are able to direct their efforts on the (program, seed) configuration most likely to lead to a desirable outcome. Attempting to enumerate the decision process in a similar way (even for a single program) by constructing tuples of (seed, mutators) leads to millions of potential arms for the bandit to choose from, again faced with the strict time budget caused by the enormous testing speed. 

Although diverse approaches exist with the goal of improving fuzzing through improved seed selection \cite{b45}, input filtering \cite{b32} and interesting new mutation operators or strategies \cite{b22}, \cite{b23}, \cite{b33}, \cite{b44}, our research provides an architecture that is complementary to them and designed with the intent to build a framework to not only enhance learning but to leverage the universal logging, tracking and benchmarking capabilities provided by OpenAI Gym and Tensorflow. Lastly, since our work is inspired by the success of the Arcade Learning Environment and ideas created for that domain, we hope to encourage further cross-pollination of ideas between RL and software testing.

\section{Threats to Validity}
The primary goal of this effort was the development of a framework to integrate existing fuzzing tools with reinforcement learning. Despite our efforts, there are still several areas that can be improved. In this section we describe some of these threats to validity with the goal of furthering research and discussion in this domain.

\subsection{Unique line coverage as a performance metric}
The presented research is focused on line coverage as an evaluation metric and not the number of faults or security vulnerabilities uncovered. While line coverage attempts to achieve the maximum breadth of code execution, it is certainly not guaranteed to effectively lead to the discovery faults.  In addition, effectively exercising rare conditions or branches \cite{b22},\cite{b44} may be crucial to fault discovery yet we do not reward this with our coverage-only metric. To summarize, reaching the dragon's lair (coverage) is not the same as slaying the dragon (finding faults or exercising a crucial branch).

The reward function described in this paper has the flexibility to be used to encourage bug discovery rather than line coverage, by rewarding the RL-Agent each time a Sanitizer (Table 2) detects an error. Unfortunately, this modification limits the class of faults that can be rewarded to those that can be detected automatically. In order to implement a "fault focused" RL agent it would also be necessary to choose how much to reward the RL algorithm for each discovered fault. 

\subsection{Limitations of Intelligent Mutator Selection}
A key goal of our experiment was to determine if we could achieve a significant improvement in fuzzing performance by only modifying the mutation operator selection process. This was done intentionally to evaluate RL in the context of simple controlled experiment. However, it is important to consider that improving mutator selection alone is likely not enough to achieve thorough testing of a target program. As discussed previously, we designed our framework to be flexible in this regard to facilitate further research in other relevant fuzzing areas such as branch targeting \cite{b33}, seed selection/filtering \cite{b32} or even to enhance the types of mutator actions being performed \cite{b22}, \cite{b23}     
\subsection{Expanding Size and Scope of Tested Programs}
It was our goal to evaluate our approach on a set of relatively well known software testing benchmarks provided by the libFuzzer test suite \cite{b25}.   However, analysis on a larger set of more complex functions would reveal additional strengths and weaknesses of our approach. In particular, further discussion of how the function or domain of software being tested affects the efficacy of our approach is still an open question. For example, why did we perform better on re2/sqlite than on boringssl? 

\section{Conclusion}
Inspired by the success of reinforcement learning in challenging, noisy domains we show that even with a limited amount of training time (about 8 hours) we are able to achieve a substantial improvement in the amount of code coverage simply by choosing mutation operators more effectively. Our approach is demonstrated across five benchmark programs evaluated under realistic conditions using substantially more runs than most existing fuzzing research. 

We demonstrate that by strategically choosing mutators we can achieve deeper coverage across all testing programs, even when the current state-of-the-art random selection is given more chances to succeed (Table 4). While this initial effort to learn how to select mutators more intelligently has been successful, this pales in comparison to what our architecture can achieve in the future, uniting large-scale fuzzing (libFuzzer) and learning (OpenAI) under a single multi-platform, multi-language distributed architecture (via GRPC). By using asynchronous buffers as a mechanism to coordinate between fuzzing and machine learning, we lift the burden that the microseconds of latency fuzzing expects, enabling in the future a system for fair evaluation and comparison between RL approaches with varying computational constraints.

\section{Acknowledgement}
I would like to thank all of the many reviewers who have helped shape the direction of the presented work.

This project was funded by the Test Resource Management Center  (TRMC)  Test  and  Evaluation  /  Science  \&  Technology  (T\&E/S\&T)  Program  through  the  U.S.  Army  Program Executive  Office  for  Simulation,  Training  and  Instrumentation  (PEO  STRI)  under  Contract  No.  W900KK-16-C-0006,“Robustness  Inside-Out  Testing  (RIOT).”  NAVAIR  Public Release 2018-165. Distribution Statement A - “Approved for public release; distribution is unlimited”.


\begin{thebibliography}{00}
\bibitem{b1}		Harold Abelson, Gerald Jay Sussman, and Julie Sussman. Structure and Interpretation of Computer Programs. MIT Press, Cambridge, Massachusetts, 1985.
\bibitem{b2}		Igor Adamski, Robert Adamski, Tomasz Grel, Adam Jędrych, Kamil Kaczmarek, Henryk Michalewski. Distributed Deep Reinforcement Learning: Learn how to play Atari games in 21 minutes. https://arxiv.org/abs/1801.02852
\bibitem{b3}		Marc G. Bellemare, Yavar Naddaf, Joel Veness, and Michael Bowling. The arcade learning environment: An evaluation platform for general agents. J. Artif. Intell. Res. (JAIR), 47:253–279, 2013.
\bibitem{b4}		M. Bohme, V.-T. Pham, and A. Roychoudhury, “Coverage-based greybox fuzzing as markov chain,” in CCS’16. New York, NY, USA: ACM, 2016, pp. 1032–1043.
\bibitem{b5}		Marcel Böhme, Van-Thuan Pham, Manh-Dung Nguyen, and Abhik Roychoudhury. 2017. Directed Greybox Fuzzing. In Proceedings of the 2017 ACM SIGSAC Conference on Computer and Communications Security (CCS ’17).
\bibitem{b6}		G. Brockman, V. Cheung, L. Pettersson, J. Schneider, J. Schulman, J. Tang, and W. Zaremba. “OpenAI Gym”.  arXiv:1606.01540 (2016).

\bibitem{b7}		Konstantin Böttinger, Patrice Godefroid, Rishabh Singh. Deep Reinforcement Fuzzing. https://arxiv.org/abs/1801.04589  - 2018

\bibitem{b8}		Sang Kil Cha, Maverick Woo, and David Brumley. 2015. Program-adaptive mutational fuzzing. In SP. 725–741.
\bibitem{b9}		Bruce P. Douglass. 1998. Statecarts in use: structured analysis and object-orientation. In Lectures on Embedded Systems,
\bibitem{b10}		M.Eddington,“2015.Peach fuzzer platform. (2015). http://www.peachfuzzer.com/products/peach-platform/.
\bibitem{b11}		Vijay Ganesh, Tim Leek, and Martin Rinard. Taint-based directed
whitebox fuzzing. In Proceedings of the 31st International Conference
on Software Engineering, pages 474–484. IEEE Computer Society, 2009.
\bibitem{b12}		Patrice Godefroid, Michael Y Levin, and David Molnar. Sage: whitebox fuzzing for security testing. Queue, 10(1):20, 2012.
\bibitem{b13}		Patrice Godefroid, Hila Peleg, and Rishabh Singh. Learn and fuzz: Machine learning for input fuzzing.” in Proceedings of the 32nd IEEE/ACM International Conference on Automated Software Engineering (ASE 2017). IEEE Press, 2017, pp. 50–59.
\bibitem{b14}		Van Hasselt, Hado, Guez, Arthur, and Silver, David. Deep reinforcement learning with double q-learning. arXiv preprint arXiv:1509.06461, 2015.
\bibitem{b15}		Guided in-process fuzzing of Chrome components https://security.googleblog.com/2016/08/guided-in-process-fuzzing-of-chrome.html
\bibitem{b16}		Matthew J. Hausknecht and Peter Stone. Deep recurrent q-learning for partially observable mdps. Proc. of Conf. on Artificial Intelligence, AAAI, 2015
\bibitem{b17}		N. Heess, S. Sriram, J. Lemmon, J. Merel, G. Wayne, Y. Tassa, T. Erez, Z. Wang, A. Eslami, M. Riedmiller, et al. “Emergence of Locomotion Behaviours in Rich Environments”arXiv  arXiv:1707.02286 (2017)
\bibitem{b18}		Ulf  Kargén and Nahid Shahmehri. Turning programs against each other: high coverage fuzz-testing using binary-code mutation and dynamic slicing. In FSE. 782–792.

\bibitem{b19}		Krakovsky, M. (2016). Reinforcement renaissance. Communications of the ACM, 59(8):12–14.

\bibitem{b20}       Nathan P Kropp, Philip J Koopman, and Daniel P Siewiorek. 1998. Automated robustness testing of off-the-shelf software components. In Proceedings of the 28th Annual International Symposium on Fault-Tolerant Computing. IEEE, 230–239.

\bibitem{b21}		LATTNER, C., AND ADVE, V LLVM: A compilation framework for lifelong program analysis and transformation. Proceedings of the international symposium on Code generation and optimization (CGO 2004).
\bibitem{b22}		Caroline Lemieux, Koushik Sen. FairFuzz: Targeting Rare Branches to Rapidly Increase Greybox Fuzz Testing Coverage 

\bibitem{b23}		Yuekang Li, Bihuan Chen, Mahinthan Chandramohan, Shang-Wei Lin, Yang Liu, and Alwen Tiu. 2017. Steelix: Program-state Based Binary Fuzzing. In Proceedings of the 2017 11th Joint Meeting on Foundations of Software Engineering (ESEC/FSE 2017)
\bibitem{b24}		LibFuzzer: A library for coverage-guided fuzz testing. http: //llvm.org/docs/LibFuzzer.html. (2017)

\bibitem{b25}		libFuzzer: Fuzzer Test Suite https://github.com/google/fuzzer-test-suite

\bibitem{b26}		Barton P. Miller, Lars Fredriksen, and Bryan So. An empirical study of the reliability of UNIX utilities. Commun. ACM, 33(12):32–44, 1990
\bibitem{b27}		V. Mnih, K. Kavukcuoglu, D. Silver, A. Graves, I. Antonoglou, D. Wierstra, et al. Playing Atari with deep reinforcement learning. Technical report Deepmind Technologies (2013) arXiv:1312.5602 

\bibitem{b28}		Volodymyr Mnih, Adria Puigdomenech Badia, Mehdi Mirza, Alex Graves, Timothy Lillicrap, Tim Harley, David Silver, and Koray Kavukcuoglu. Asynchronous methods for deep reinforcement learning. In International Conference on Machine Learning, 1928–1937, 2016.
\bibitem{b29}		Nicole Nichols, Mark Raugas, Robert Jasper, Nathan Hilliard. Faster Fuzzing: Reinitialization with Deep Neural Models  https://arxiv.org/abs/1711.02807v1

\bibitem{b30}		OSS-Fuzz: Five months later, and rewarding projects https://opensource.googleblog.com/2017/05/oss-fuzz-five-months-later-and.html

\bibitem{b31}		Van-Thuan Pham, Marcel Böhme, and Abhik Roychoudhury. 2016. Model-based whitebox fuzzing for program binaries. In ASE. 543–553.
\bibitem{b32}		Mohit Rajpal, William Blum, Rishabh Singh. Not all bytes are equal: Neural byte sieve for fuzzing. https://arxiv.org/abs/1711.04596., 2017.

\bibitem{b33}		Sanjay Rawat, Vivek Jain, Ashish Kumar, Lucian Cojocar, Cristiano Giuffrida, and Herbert Bos. 2017. VUzzer: Application-aware Evolutionary Fuzzing. In NDSS

\bibitem{b34}		A. Rebert, S. K. Cha, T. Avgerinos, J. Foote, D. Warren, G. Grieco, and D. Brumley, “Optimizing seed selection for fuzzing,” in USENIX Security, 2014, pp. 861–875

\bibitem{b35}		Schaarschmidt, Michael and Kuhnle, Alexander and Fricke, Kai. Tensorforce Benchmark CartPole DDQN \url{https://github.com/reinforceio/tensorforce/benchmark/blob/master/configs/ddqn_cartpole}

\bibitem{b36}		Schaarschmidt, Michael and Kuhnle, Alexander and Fricke, Kai. TensorForce: A TensorFlow library for applied reinforcement learning. https://github.com/reinforceio/tensorforce

\bibitem{b37}		Guy Eric Schalnat, Andreas Dilger, John Bowler, and Glenn et. al. Randers-Pehrson. readpng and libpng. \url{http://www.libpng.org/pub/png/}

\bibitem{b38}		Schaul, Tom, Quan, John, Antonoglou, Ioannis, and Silver, David. Prioritized experience replay. arXiv preprint arXiv:1511.05952, 2015
\bibitem{b39}		Kostya Serebryany. OSS-Fuzz - Google's continuous fuzzing service for open source software.USENIX 2017
\bibitem{b40}		Konstantin Serebryany, Derek Bruening, Alexander Potapenko, and Dmitry Vyukov. 2012. AddressSanitizer: a fast address sanity checker. In Proceedings of the 2012 USENIX conference on Annual Technical Conference (USENIX ATC'12). USENIX Assoc. Berkeley, CA, USA, 28-28
\bibitem{b41}		David Silver, Julian Schrittwieser, Karen Simonyan, Ioannis Antonoglou, Aja Huang, Arthur Guez, Thomas Hubert, Lucas Baker, Matthew Lai, Adrian Bolton, Yutian Chen, Timothy Lillicrap, Fan Hui, Laurent Sifre, George van den Driessche, Thore Graepel, and Demis Hassabis. Mastering the game of go without human knowledge. Nature, 550:354– 359, 2017.
\bibitem{b42}		S. Shah, D. Sundmark, B. Lindstrm, and S. F. Andler, “Robustness Testing of Embedded Software Systems: An Industrial Interview Study,” IEEE Access, vol. 4, pp. 1859–1871, 2016
\bibitem{b43}		 Evgeniy Stepanov, Konstantin Serebryany. MemorySanitizer: fast detector of uninitialized memory use in C++. Proceedings of the 2015 IEEE/ACM International Symposium on Code Generation and Optimization (CGO), CGO 2015, San Francisco, CA, USA, pp. 46-55 
\bibitem{b44}		Nick Stephens, John Grosen, Christopher Salls, Andrew Dutcher, Ruoyu Wang, Jacopo Corbetta, Yan Shoshitaishvili, Christopher Kruegel, and Giovanni Vigna. 2016. Driller: Augmenting Fuzzing Through Selective Symbolic Execution. In NDSS.

\bibitem{b45}		Junjie Wang, Bihuan Chen, Lei Wei, and Yang Liu. 2017. Skyfire: Data-Driven Seed Generation for Fuzzing. In SP. 579–594.

\bibitem{b46}		Maverick Woo, Sang Kil Cha, Samantha Gottlieb, and David Brumley Scheduling black-box mutational fuzzing. In CCS. 511–522.
\bibitem{b47}		Michał Zalewski. American fuzzy lop. \url{http://lcamtuf.coredump.cx/afl/}
\end{thebibliography}
\end{document}